\documentclass[%
 aip,
 reprint,%
]{revtex4-1}

\usepackage{graphicx}
\usepackage{dcolumn}
\usepackage{bm}

\usepackage[utf8]{inputenc}
\usepackage[T1]{fontenc}
\usepackage[dvipsnames]{xcolor}
\usepackage{amsmath}
\usepackage{amssymb}
\usepackage{mathptmx}

\newcommand{\rev}[1]{\textcolor{black}{#1}}

\begin{document}

\preprint{AIP/123-QED}

\title{Progress and Challenges in \rev{Ab Initio} Simulations of Quantum Nuclei \rev{in Weakly Bonded Systems}}

\author{Mariana Rossi}
\affiliation{ 
Max Planck Institute for the Structure and Dynamics of Matter, Luruper Chaussee 149, 22761 Hamburg, Germany 
}%

\date{\today}

\begin{abstract}
Atomistic simulations based on the first-principles of quantum mechanics are reaching unprecedented length scales. This progress is due to the growth in computational power allied with the development of new methodologies that allow the treatment of electrons and nuclei as quantum particles. In the realm of materials science, where the quest for desirable emergent properties relies increasingly on soft weakly-bonded materials, such methods have become indispensable. In this perspective, an overview of simulation methods that are applicable for large system sizes and that can capture the quantum nature of electrons and nuclei in the adiabatic approximation is given. In addition, the remaining challenges are discussed, especially regarding the inclusion of nuclear quantum effects (NQE) beyond a harmonic or perturbative treatment, the impact of NQE on electronic properties of weakly-bonded systems, and how different first-principles potential energy surfaces can change the impact of NQE on the atomic structure and dynamics of weakly bonded systems. 
\end{abstract}

\maketitle


\section{\label{sec:intro} Introduction}

Quantum mechanical simulations with atomistic resolution are currently reaching length and time scales that had been almost unimaginable only a few decades ago. The reasons for this impressive progress can be attributed to a number of conspiring factors including the increase of the availability of high performance computers, improved algorithms in many community software packages \cite{Yu:2019to, Oliveira:2020uz, Kapil:2019ju}, and the sympatico relationship of machine learning with quantum mechanical methods \cite{JinnouchiKresseBokdam_prl_2019, JiaLinCarZhang_arxiv_2020, Buxton:2017ck, Chmiela:2019bg}. 

\begin{figure}[ht]
    \centering
    \includegraphics[width=0.32\textwidth]{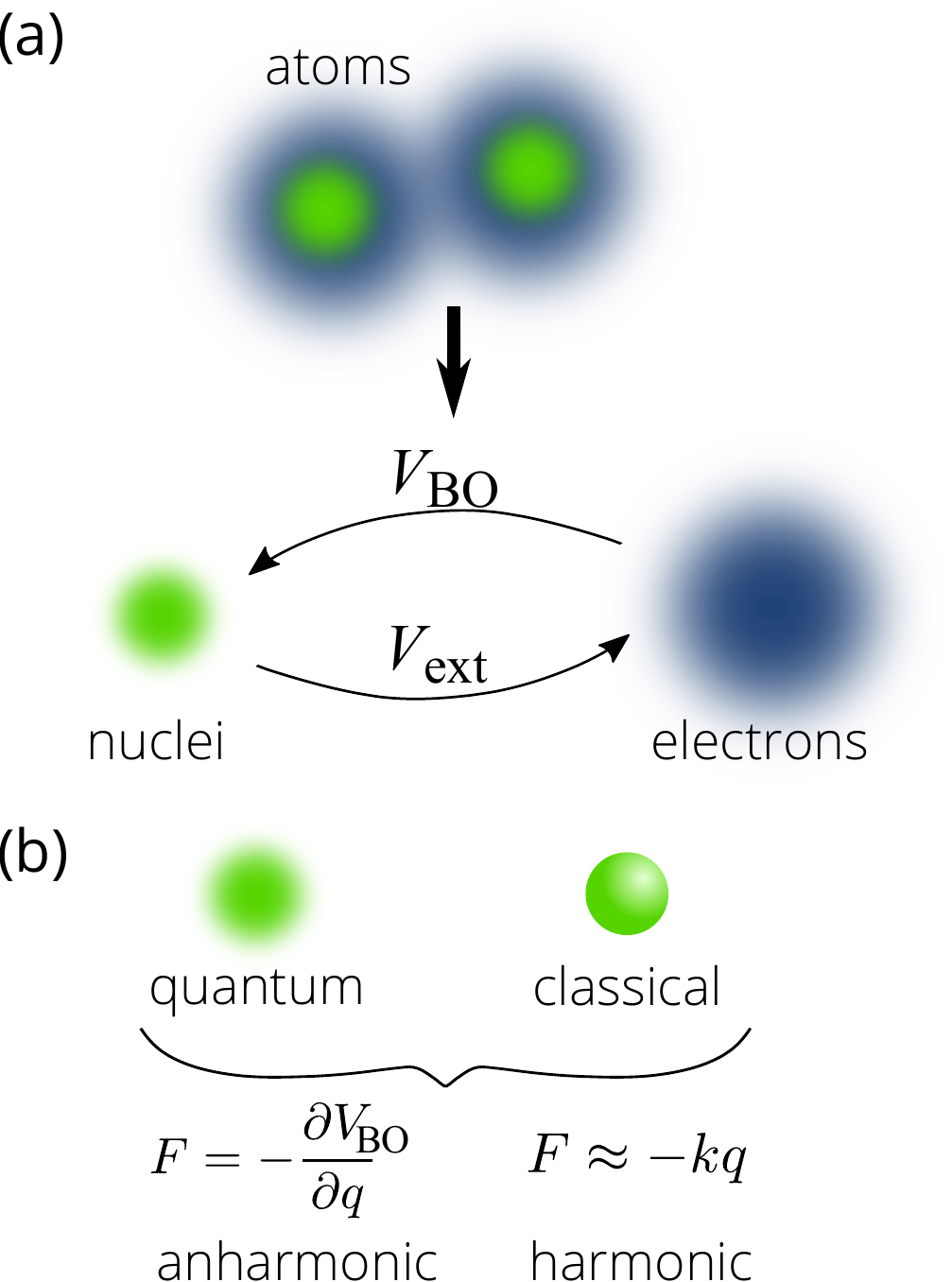}
    \caption{A sketch depicting the conventional model of atoms under the Born-Oppenheimer approximation in (a). A nuclear configuration provides the commonly called ``external potential" ($V_{ext}$) that the electrons are subject to, while the ground state expectation value of the electronic Hamiltonian provides the Born Oppenheimer potential ($V_{BO}$), which enters as the potential energy of the nuclear degrees of freedom. Under this approximation, it is possible to treat the nuclei as quantum or classical particles, as shown in (b). In turn, in both cases, one can assume that the forces acting on the nuclei are approximately harmonic, or not, being described by the gradient with respect to nuclear positions $q$ of the Born Oppenheimer potential.}
    \label{fig:fig1}
\end{figure}

The fundamental theoretical basis for the majority of the atomistic simulations that can be found in the literature to date is the celebrated Born-Oppenheimer approximation. Within this approximation, the quantum dynamics of electrons and nuclei can be treated separately, and one is free to employ further approximations either on the electronic or on the nuclear problem, as sketched in Fig. \ref{fig:fig1}(a).

The need for a quantum mechanical treatment of electrons has long been well accepted\cite{Bohr1913}. Despite this fact, classical-like functional forms (or, force fields) that do not explicitly treat the electronic degrees of freedom individually but rather as an integral part of an effective atom, can still offer the best compromise between cost and accuracy for simulations involving system sizes beyond a few thousand atoms or time scales reaching beyond the nanosecond regime~\cite{Shaw2010}. These force fields can often deliver good qualitative insights, but they typically lack broad transferability and quantum mechanical accuracy. Treating the electronic degrees of freedom with ``first-principles'' methods restores the transferability, while improving the accuracy, of the overall approach.

On the other hand, the assumption that the nuclei behave classically in most situations has been left essentially unchallenged for decades \rev{in many areas of physical chemistry and materials science, with few notable exceptions \cite{ParrinelloRahman1984, Marx:1995ch, MarxParrinello1995, Poulsen:2003fy}}. However, as experimental measurements become more precise (e.g., single molecule or time resolved ultrafast experiments) and theoretical methods that can treat  systems on a full quantum mechanical picture become more computationally feasible, examples spanning all classes of  materials start to emerge,
where the quantum nature of the nuclei provokes a qualitative change in several quantities~\cite{Markland:2018fv}, even in the absence of electronically non-adiabatic transitions. 

In \rev{some} situations, the quantum nature of the nuclei has been acknowledged: The addition of zero-point energy corrections within the harmonic approximation is a standard practice,
for example, when analysing relative and reaction energies in the gas-phase and in some solid-state problems. However, treating nuclei as quantum particles still represents a challenge in a number of \rev{ways}, including (i) going beyond the harmonic approximation, (ii) addressing nuclear dynamics, and (iii) addressing soft materials and liquids. For the most part, these calculations would require quantum nuclei with anharmonic forces (see Fig.~\ref{fig:fig1}(b)). \rev{Taking these effects into account can lead to a proper treatment of electron-phonon coupling, which has several implications for superconductivity, metal-insulator transitions, charge transfer and transport, and others.}

This perspective aims to provide an overview of  methodology that can be employed to treat both electrons and nuclei as quantum particles in large-scale atomistic simulations, with a focus on the field of \textit{ab initio} path integral molecular dynamics. \rev{It does not cover any techniques going beyond the adiabatic approximation for electronic and nuclear degrees of freedom. The perspective }includes considerations to estimate whether NQE are relevant for a specific problem, and how the shape of the potential energy surface can influence the degree to which NQE lead to the observed structures and dynamics of the system. Situations where the use of perturbative approaches to include NQE may not be sufficient are shown and the impact of these nuclear fluctuations on the electronic structure of weakly bonded systems is discussed.

\section{\label{sec:theoey-elec} Quantum mechanics for electrons in large scale simulations}

Under the Born-Oppenheimer approximation, the electronic structure problem is solved at a fixed configuration of the nuclei, which defines the external potential $V_{ext}$, and the electrons create the Born-Oppenheimer (BO) potential $V_{BO}$ \rev{that defines the potential energy surface} where the nuclei move. \rev{Accurately capturing potential energy surfaces, polarization, electronic density rearrangement, etc., is essential for a reliable assessment of the impact of NQE on diverse properties of materials.}

The \rev{BO potentia canl} be obtained with an electronic structure method of choice, for example using wave-function based methods or density-functional theory (DFT). Such methods, which consider electrons as explicit quantum particles, are very accessible and tractable nowadays. A recent special issue of the Journal of Chemical Physics~\cite{JCP-ES} highlights the advances in many electronic structure packages, showing how algorithms to solve electronic structure and dynamics problems have evolved together with computer architectures in the last decades. Because of these algorithmic developments, very accurate methodologies, like coupled cluster theory, have become available to treat larger systems including periodic solids~\cite{RiplingerNeeseJCP,ZhangGruneisCCSDT}. 

Within the realm of DFT, one can obtain very reliable results by employing good approximations to the exchange-correlation potential, $\nu_{xc}=\delta E_{xc}[\rho]/\delta \rho$, where $\rho$ is the electronic density. Even though all known approximations to $\nu_{xc}$ are not exact, they yield, in many cases, reliable electronic properties for molecules, liquids and solids. In addition, they can also provide an accurate potential energy surface, from which structural and dynamical properties of the nuclei can be \rev{calculated~\cite{Rossi:2010exa,Rossi:2014gq, Raimbault:2019ex, Marsalek:2017kn,Li:2011fd,LitmanJACS2019}}, 
which attests to their capacity to deliver reliable results for diverse nuclear configurations. In particular, approximations that take into account fractions of exact exchange in this term mitigate the self-interaction error~\cite{Cohen:2011fm}, and range-separated functionals can, moreover, approximate electronic screening~\cite{Krukau:2008it}. This family of so-called ``hybrid'' functionals is more computationally demanding, but recent implementations of these methods are very efficient, and allow the treatment of either thousands of atoms~\cite{Janke:2020ff} or thousands of force evaluations for smaller systems~\cite{Cheng:tz}.
Tailor-made and local hybrid functionals capable of delivering good accuracy for interfaces have also been recently developed~\cite{Borlido:2018jm,Zheng:2019fo}. 

\rev{For weakly bonded systems, capturing long range dispersion interactions (usually simply called van der Waals interactions) is of extreme importance \cite{JohnsonYang2010, Marom:2011vdW, RossiTkaVarma2013, TkatchenkoRossi2011, FidanyanRossi2020}. These interactions are highly non-local and thus absent from standard DFT functionals, due to their inherent spatial locality. They can be accounted by non-local functionals~\cite{Berland:2015kj}, or by corrections that are added to semilocal (and hybrid) functionals~\cite{BeckeJohnson2005, Tkatchenko:2009TS, GrimmeD32010, Gould:2016jn, Tkatchenko:2012MBD}. With these functionals and corrections, dispersion interactions can be accurately captured in a wide range of systems, spanning from condensed phase solids and surfaces, to soft molecular matter. So far, no correction or functional in this area was shown to perform equally well across several different material classes, which still presents a challenge especially for new material architectures.}

\rev{There are still a number of open challenges to be tackled within electronic structure theory, in general, and within DFT in particular~\cite{Cohen:2011fm}. The lack of an exact density functional means that a very large amount of exchange-correlation approximations exist and it is often difficult to evaluate whether a particular one can yield predictive results for a given (large-scale) problem. Employing approximations that can capture the relevant physics of the problem at hand guarantees a certain degree of reliability. However, without the availability of benchmarks from high-level quantum-chemistry methods or appropriate experiments, it is usually not possible to ascertain the accuracy of DFT approximations for certain properties of complex systems. Nevertheless, the quality of the current DFT approximations and the possibility of applying many-body perturbation methods~\cite{Reining:2018ek} or quantum-chemistry wave-function methods to larger systems has led to a situation where a proper theoretical treatment of the nuclear problem becomes necessary to unravel further physical phenomena that can be assessed in modern experiments.}

\section{Quantum mechanics for nuclei in large scale simulations \label{sec:theory}}

\begin{figure}[ht]
    \centering
    \includegraphics[width=0.48\textwidth]{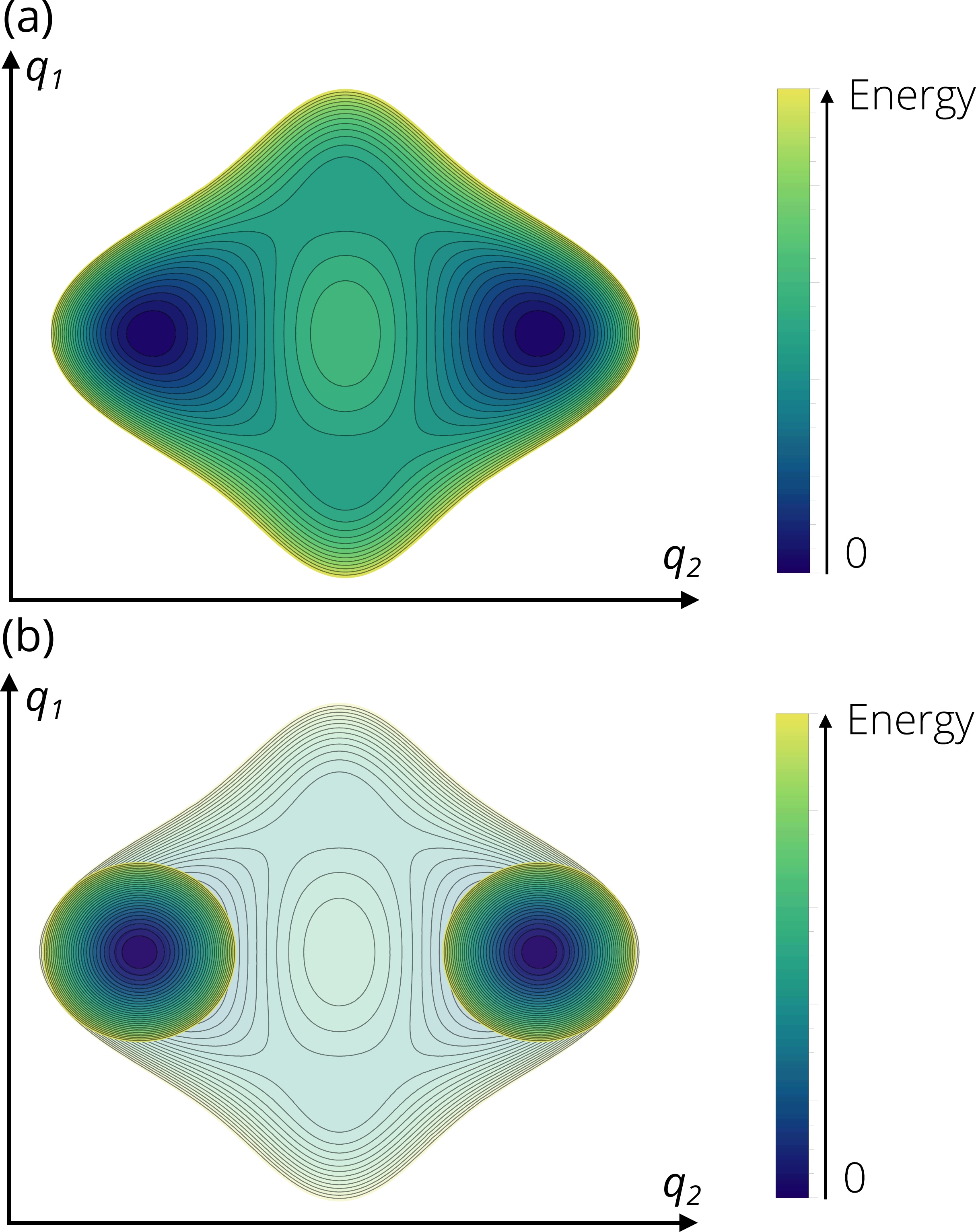}
    \caption{(a) Example of a two-dimensional potential energy surface represented by a quartic potential along $q_1$ and $q_2$ with quadratic coupling between the coordinates. (b) Corresponding harmonic approximation of the potential energy surface depicted in (a) around \rev{its two lowest energy minima} in solid colors, overlaid on the original surface. \rev{The approximation distorts the potential around the minima, completely neglecting the coupling between the two coordinates}.}
    \label{fig:fig2}
\end{figure}

Regarding the nuclei, the simplest and most straightforward way to account for their quantum nature on first-principles potential energy surfaces is the so-called harmonic approximation. A quantum-mechanical Hamiltonian for the nuclei is proposed in which the potential is considered to be of a quadratic form with respect to the nuclear coordinates. This is achieved by a truncation after the second order of a Taylor expansion of the potential around a minimum,
\begin{equation}
    V_{\text{BO}}(\bm{q}) \approx V_{\text{BO}}(\bm{q}_0)+\frac{1}{2}\sum_{i,j}^{3N} q_i \left.\frac{\partial V_{\text{BO}}(\bm{q})}{\partial q_i \partial q_j}\right\rvert_{0} q_j  \label{eq:harmonic},
\end{equation}
where $q_i$ are nuclear coordinates, $N$ is the number of atoms in the system and $0$ denotes a global or local minimum of the BO potential. The second term on the right involves the second derivative of the potential with respect to nuclear positions (the Hessian) and can be written in diagonal form by a coordinate transformation. 
The solution of such a \rev{Schr\"odinger} equation for the nuclei yields the nuclear vibrational modes and their respective energies. These solutions lead to several analytical results for diverse quantities, as for example the evaluation of zero point energy, free energies, vibrational spectra, reaction rates, and others~\cite{Wilson:2158285,McQuarrie:102388}. In particular, for periodic solids, such a formulation presents the advantage of easily accounting for the phonon dispersion within the Brillouin zone, by considering the periodicity of atomic displacements and their phases in reciprocal space~\cite{Ashcroft:102652}.
However, what this picture completely disregards is the coupling between different vibrational modes, as well as anharmonic terms pertaining to any given mode. In Figure~\ref{fig:fig2}(a,b) a pictorial representation of a harmonic approximation to a considerably anharmonic 2-dimensional double-well potential is shown. In this potential, it is straightforward to visualize how the coupling between two coordinates is neglected with this approximation, and the anharmonic terms around the chosen minimum are also absent. 

There are different routes to include anharmonic contributions in the nuclear motion. Keeping with a quantum picture of nuclei, one route is to propose harmonic potentials that effectively capture anharmonic terms\rev{~\cite{Hellman:2011gi,Brown:2013gu,Errea:2014ke,Kapil:2019ch}}. Another is to consider higher-order terms in a perturbative Taylor series expansion of the potential\rev{~\cite{YuBowman2015,Konig:2020ek, Monserrat:2013ct, Norris:1998bz, Kapil:2019ch,Fortenberry2021}} or, for certain quantities, simply include anharmonic corrections from fitted potential forms to  vibrational modes that are particularly anharmonic~\cite{Hoja:2019hw}. 
However, the cost of the more rigorous theories tends to have a very steep scaling with the number of degrees of freedom, making them feasible only for \rev{small systems or for larger systems if a small subset of essential degrees of freedom are sufficient.}
In general, such theories in their most widely implemented forms, are \rev{only} applicable for soft matter and liquids \rev{where approximations can be found. The challenges in these cases involve the existence of multiple quasi-degenerate local minima, the lack of a well-defined reference configuration, and situations where} anharmonic terms cannot be treated as just a small perturbation. 

In contrast, a complete anharmonic picture can be obtained by statistically sampling the full BO potential. This can be achieved through \textit{ab initio} molecular dynamics~\cite{CarParrinello1985,Kuhne:2014bf,Niklasson:2017bj}, but with a caveat: evolving the nuclei in time (even if only for sampling purposes) supposes they follow Newtonian equations of motion generated by a classical Hamiltonian and thus behave as classical particles. 
Instead, the method of \textit{ab initio} path integral molecular dynamics~\cite{ParrinelloRahman1984,MarxParrinello1995}, which was pioneered in the 80s and has gained broad popularity in the past decade, allows one to achieve a quantum mechanical description of both nuclei and electrons, within the Born-Oppenheimer approximation.
This method delivers quantum statistics for the nuclear degrees of freedom by sampling the potential of a special classical system, making use of the ``quantum-classical isomorphism''~\cite{Feynman:100771}. Under the assumption of distinguishable particles and the BO approximation~\cite{marx_hutter_2009}, the quantum partition function in the position representation within the (\textit{ab initio}) path integral formalism can be written as
\begin{equation}
\begin{split}
Z = \lim_{P \to \infty} Z_P = \lim_{P \to \infty} \prod_{i=1}^{3N}  \left(\frac{m^{(i)}}{2 \pi \hbar^2 \beta_P}\right)^{3P/2} \int \prod_{i=1}^{3N}dq_1 \dots dq_{P} \times  \\ 
\mathrm{Exp}\left\{ -\beta_P \left[\sum_{i=1}^{3N} \sum_{j=1}^{P} \frac{m^{(i)} \omega_P^2}{2} (q^{(i)}_{j+1}-q^{(i)}_{j})^2  +  
\sum_{j=1}^{P} V_{\text{BO}}(q^{(1)}_j, \dots, q^{(3N)}_j) \right] \right\}
\end{split}
\end{equation}
where $P$ is a convergence parameter that corresponds to the discretization of imaginary time, or the commonly called ``beads'' of a ring polymer, $\beta_P=1/(P k_B T)$, $\omega_P=Pk_BT/\hbar$, and $q_j^{(i)}$ represents the position degree of freedom $i$ in the $j$th bead of the ring-polymer. In practice, these imaginary-time slices, or beads, represent replicas of the full physical system of interest, which are connected to each other through the harmonic spring terms. The reintroduction of a kinetic energy term in the exponential allows the formulation of a Hamiltonian that can be sampled through molecular dynamics in the extended phase-space of the ring polymer~\cite{ParrinelloRahman1984}, namely
\begin{eqnarray}
H_P = \sum_{j=1}^P \sum_{i=1}^{3N} \left[ \frac{(p_j^{(i)})^2}{2m'^{(i)}} + \frac{1}{2}m^{(i)} \omega_P^2 (q_j^{(i)} - q_{j-1}^{(i)})^2 \right] \nonumber \\ + \sum_{j=1}^P  V_{\text{BO}}(q_j^{(1)}, q_j^{(2)}, \dots, q_j^{(3N)})\label{eq:rp-ham}.
\end{eqnarray}
When $P$ tends to infinity, sampling configurations with the equations of motion generated by $H_P$ gives access to the quantum Boltzmann distribution of particles in the system, and hence guarantees the correct quantum statistics in the limit of distinguishable particles. The masses $m'$ entering the kinetic energy term in Eq. \ref{eq:rp-ham} can assume any value as they only serve to assist with sampling.

Unlike \textit{ab initio} molecular dynamics with classical nuclei, where the dynamics algorithm is both a tool for statistical sampling and a means of real time propagation of nuclear coordinates, in path integral molecular dynamics the time evolution of the equations of motion serves only as a sampling tool. 
It is possible to formulate real-time path integral methods to access the quantum real-time evolution of nuclei. However, even with notable progresses being achieved in the last years~\cite{Makri:2018jm}, the applicability of these methods remains rather limited, due to the sign problem arising from the oscillatory nature of the real-time propagator. Semiclassical methods, based on different semiclassical approximations of the real-time propagator, can  overcome the sign problem~\cite{Miller2001}. In particular, the ring-polymer instanton method~\cite{Richardson:2018ei}, which uses trajectories in imaginary time to describe nuclear tunnelling along well-defined reaction coordinates, has been successfully applied to quite high-dimensional systems, including molecules adsorbed on metallic surfaces~\cite{LitmanRossi2020,Fang:2020ky}.  

Attempts to approximate quantum dynamics with techniques based on imaginary-time path integrals \cite{CaoCMD1994, CraigRPMD2004, RossiTRPMD2014, LiuPILD2016} can also circumvent the sign problem, and thus dramatically extend their applicability realm -- but these approximations to quantum dynamics are often not mathematically justified.
Nevertheless, these methods have met considerable success in the simulation of vibrational spectroscopy, diffusion properties, reaction rates, and others~\cite{LiuPILD2016, Medders:2016gz, Habershon:2013kc, LitmanJACS2019, Marsalek:2017kn, RossiDiff2016, Cendagorta:2016if, Jung:2020il}. Thermostatted ring polymer molecular dynamics (TRPMD)~\cite{RossiTRPMD2014} is a very efficient approximation in this area, which makes it useful when working with costly \textit{ab initio} potentials. This approximation boils down to the dynamics in the extended phase space of the ring polymer (with $m'^{(i)} = m^{(i)}$), as proposed in ring polymer molecular dynamics (RPMD)~\cite{CraigRPMD2004}, but with thermostats attached only to the internal modes of the ring polymer. The approximation to Kubo-transformed correlation functions $\bar{c}_{AB}$ in RPMD/TRPMD is given by
\begin{equation}
\begin{split}
    \bar{c}_{AB}(t) \approx \lim_{P\to\infty} \frac{1}{(2 \pi \hbar)^P Z_P} \int d\bm{q}_1 \dots d\bm{q}_P d\bm{p}_1\dots d\bm{p}_P \times \\ A_P(\bm{q}(0))B_P(\bm{q}(t)) e^{-\beta_P H_P} 
\end{split}
\end{equation}
where $A_P(\bm{q}) = \sum_{j=1}^P A(\bm{q}_j)/P$ and $\bm{q}$ ($\bm{p}$) is the vector of all $3N$ Cartesian positions (momenta) of the system. 
The time evolution of position and momenta in TRPMD is given by
\begin{eqnarray}
     \frac{\partial q^{(i)}_j}{\partial t} &=& p^{(i)}_j/m^{(i)} \\
     \frac{\partial p^{(i)}_j}{\partial t} &=& -\nabla_i V_{BO}(q^{(1)}_j \dots q^{(3N)}_j) \nonumber\\ 
    & &- \sum_{j'} m^{(i)} \omega_P^2(2\delta_{j,j'} -\delta_{j,j'-1}-\delta_{j, j'+1})q_{j'}^{(i)} \nonumber\\
    & &- \sum_{j'} \gamma_{jj'}p_{j'}^{(i)}+\sqrt{\frac{2m^{(i)}}{\beta_P}} \sum_{j'} (\bm{\gamma}^{\frac{1}{2}})_{jj'}\xi_{j'}^{(i)}
\end{eqnarray}
where $\xi_k(t)$ is a Gaussian-distributed random number with $\langle \xi_k(t) \rangle = 0$ and $\langle \xi_k(0) \xi_k(t) \rangle = \delta(t)$ and $\bm{\gamma}$ is a friction matrix determined in the following way. We define the matrix $\bm{K}$ with elements $K_{jj'}=\omega_P^2(2\delta_{j,j'} -\delta_{j,j'-1}-\delta_{j, j'+1})$ and its transformation to the internal modes of the free ring polymer $\tilde{\bm{K}}=\bm{C}^T\bm{K}\bm{C}=\tilde{\bm{\omega}}^2$, where the transformation matrices $\bm{C}$ are given in Ref.\cite{Ceriotti_2010} and $\tilde{\omega}_{kk'}=2\omega_P \sin(k \pi / P) \delta_{kk'}$. The original scheme proposes $\tilde{\bm{\gamma}}=\tilde{\bm{\omega}}$ ($\bm{\gamma}=\bm{C}\bm{\tilde{\gamma}}\bm{C}^T$), which ensures optimal damping of the potential. In addition, $\tilde{\gamma}_{00}=0$ for the centroid mode, which ensures that the formalism maintains all quantum-mechanical limits that RPMD autocorrelation functions were shown to obey~\cite{RossiTRPMD2014}. Because the formalism imposes only a few restrictions on the friction coefficients, other white-noise friction parameters inspired by harmonic-well problems have been proposed~\cite{HeleTRPMD2016}, and the extension of the formalism to generalized Langevin equation (GLE) thermostats has also been explored~\cite{RossiTRPMD22018}. The approach is applicable, in particular, to vibrational spectra of condensed phase systems~\cite{Rossi:2014gq, Marsalek:2017kn} and molecules with many degrees of freedom~\cite{LitmanJACS2019, Litman:2019ix} at moderate temperatures. It is free from the spurious artefacts of RPMD and centroid molecular dynamics on vibrational spectra~\cite{Witt:2009hy}, albeit introducing an unphysical broadening to the lineshape of the peaks~\cite{RossiTRPMD2014}. This broadening could be partially mitigated with GLE thermostats~\cite{RossiTRPMD22018}. At a higher simulation cost, path integral Liouville dynamics can deliver excellent results for the vibrational spectra of small molecules~\cite{LiuPILD2016}.

The successes of these approximate methods rely on their exact treatment of quantum Boltzmann statistics \rev{(i.e. the quantum statistics of distinguishable particles)} for the initial conditions and their conservation of the quantum Boltzmann distribution along different types of time evolution based on classical trajectories. Their drawbacks stem from their lack of any quantum-mechanical phase information, and the fact that their largely \textit{ad hoc} nature makes it difficult to construct systematic improvements.
An interesting recent development on this front has been the derivation of an approximate quantum dynamics technique called Matsubara dynamics~\cite{HeleMATSUBARA2015}, in which quantum statistics and classical trajectories can be combined for real time properties through a rigorous route starting from first-principles quantum dynamics. Even though this method also suffers from the sign problem (rendering it impractical), it is possible to see how methods like RPMD, TRPMD and CMD can be obtained from it~\cite{HeleMATSCON2015,HeleTRPMD2016}. This theory has already opened the door to the improvement of some of these approximate methods~\cite{Trenins:2019go}, and offers potential avenues for further developments.

\section{When are nuclear quantum effects important?}

NQE on equilibrium and response properties are more pronounced at lower temperatures and for lighter particles. A quantity that illustrates this particularly well is the thermal de Broglie wavelength, given by
\begin{equation}
    \lambda = \left( \frac{2 \pi \hbar^2}{m k_B T} \right)^{\frac{1}{2}}.
\end{equation}
Even though this expression assumes non-interacting free particles, it tends to be a good estimator of the importance of NQE. Classical Boltzmann statistics typically break down if $\lambda$ is of the same order of, or larger than, the typical interparticle spacing. As an example, for a proton at room temperature $\lambda^H=1.00$ \AA, while for a heavy nucleus like Cu, $\lambda^{Cu}=0.12$ \AA. However, at a much lower temperature like $10$ K, these values are  $\lambda^H=5.50$ \AA~ and $\lambda^{Cu}=0.69$ \AA.

One telltale sign of the importance of NQE is the appearance of geometric isotope effects. 
Classically the mass of an atom cannot change static structural properties of materials, but within quantum mechanics it can. However, the magnitude of these changes depend on the character of the potential energy surface. If the local potential energy is approximately harmonic, the effect will be negligibly small. However, if it is not, the effect can be large. 

\begin{figure}
    \centering
    \includegraphics[width=0.4\textwidth]{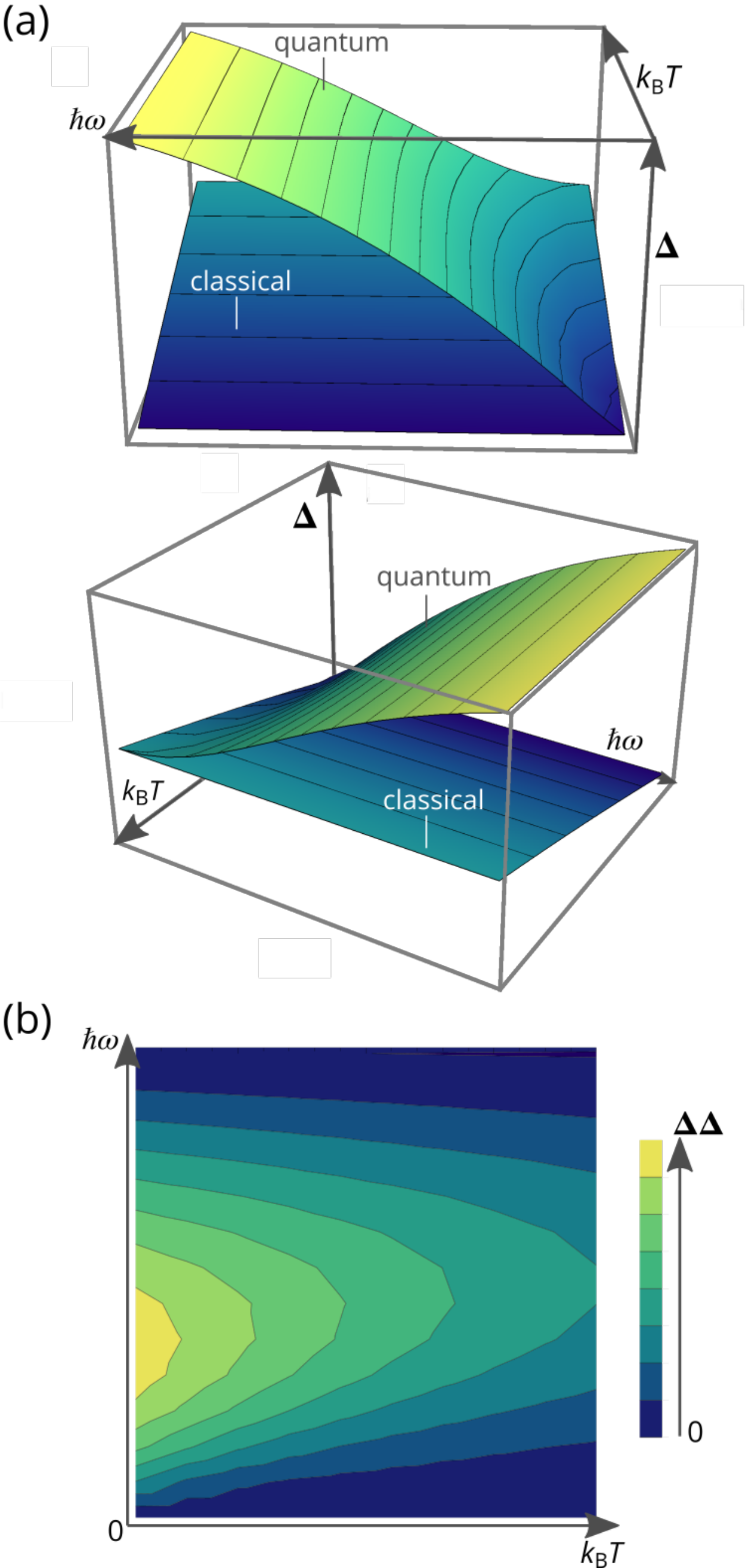}
    \caption{\rev{An example} of different regimes of nuclear fluctuations. \rev{(a) The anharmonic contribution to the quantum and classical nuclear fluctuations as a function of temperature and oscillation frequency, defined as $\Delta (T, \omega) = 1 - \langle q^2 (T, \omega) \rangle^{\text{harm.}}/\langle q^2 (T, \omega) \rangle^{\text{anh.}}$. The anharmonic expectation values were calculated based on a quartic polynomial expansion of a 1D Morse potential $V(q) = D(1-e^{-\alpha q})^2$ with $\alpha^2 =m \omega^2/(2 D)$. Parameters for the skecth were $m=m_p, D=0.1$ Ha. Two viewpoints of the same plot are presented. The region of coincidence of quantum and classical anharmonic contributions on the frequency axis increases with increasing temperature. Intervals shown in the plot span roughly up to 3500 cm$^{-1}$ and up to 600 K. (b) The difference $\Delta\Delta(T, \omega)=\Delta^{\text{qt}}(T, \omega) - \Delta^{\text{cl}}(T, \omega)$, where the anharmonic potential of a) was modified by a scaling term $(1-\omega/\omega_{max})$ that multiplies the third and fourth orders of the polynomial expansion (potential becomes harmonic approaching $\omega=\omega_{max}$). Only a region at intermediate frequencies and lower/intermediate temperatures exhibits higher degrees of ``quantum anharmonicity''.}}
    \label{fig:fig2-next}
\end{figure}

Nevertheless, even classically, the atomic mass changes dynamical properties of materials like diffusion and transport coefficients or vibrational spectra. In a harmonic potential, quantum and classical mechanics would predict the same mass-scalings for vibrational frequencies, but the occupation of the vibrational states would differ. A useful quantity to analyze for these properties is the ratio between the harmonic estimate of the zero point energy of vibrational states and their thermal occupation,  $\hbar \omega/(k_B T)$, which, if much larger than $1$ hints at the (possibly large) impact of NQE. Further, if the potential is strongly anharmonic, the dependence of the vibrational energies on the mass of the atoms can be very different in quantum and classical mechanics~\cite{Litman:2019ix}. 

Beyond effects related to the zero-point-energy, tunneling is perhaps the most pronounced NQE in high-dimensional systems at finite temperatures. Especially in chemical reactions, a good estimator for the importance of tunneling is the crossover temperature $T_c$, below which nuclear tunneling will be pronounced. Assuming a parabolic barrier, $T_c=\hbar \omega^\dagger/(2 \pi k_B)$\cite{Gillan}, where $\omega^\dagger$ is the absolute value of the imaginary frequency at the transition state geometry. Deviations from a parabolic barrier, however, leads to cases where this crossover temperature ceases to be a good estimate for tunneling contributions \cite{AlvarezBarcia:2013ju}.

One could continue listing a few of these ``rules-of-thumb'', like the estimation of the Debye temperature for heat conduction studies, and others. However, what is clear is that these estimations are based on predictions that employ harmonic potentials, as results can be obtained analytically. Indeed, NQE are relatively easy to account for when most degrees of freedom behave harmonically or only exhibit small deviations from harmonic behavior. This fact underlies a number of long-standing successes in the field solid-state physics~\cite{Giustino:2017ge, Monserrat:2013ct}. Challenges arise, however, when the anharmonic nature of potential energy surfaces is essential, and especially when systems display complex electronic structural rearrangements that require an \textit{ab initio} treatment.

\section{Beyond the harmonic approximation with quantum electrons and nuclei: Examples \label{sec:examples}}

\begin{figure*}[ht]
    \centering
    \includegraphics[width=0.98\textwidth]{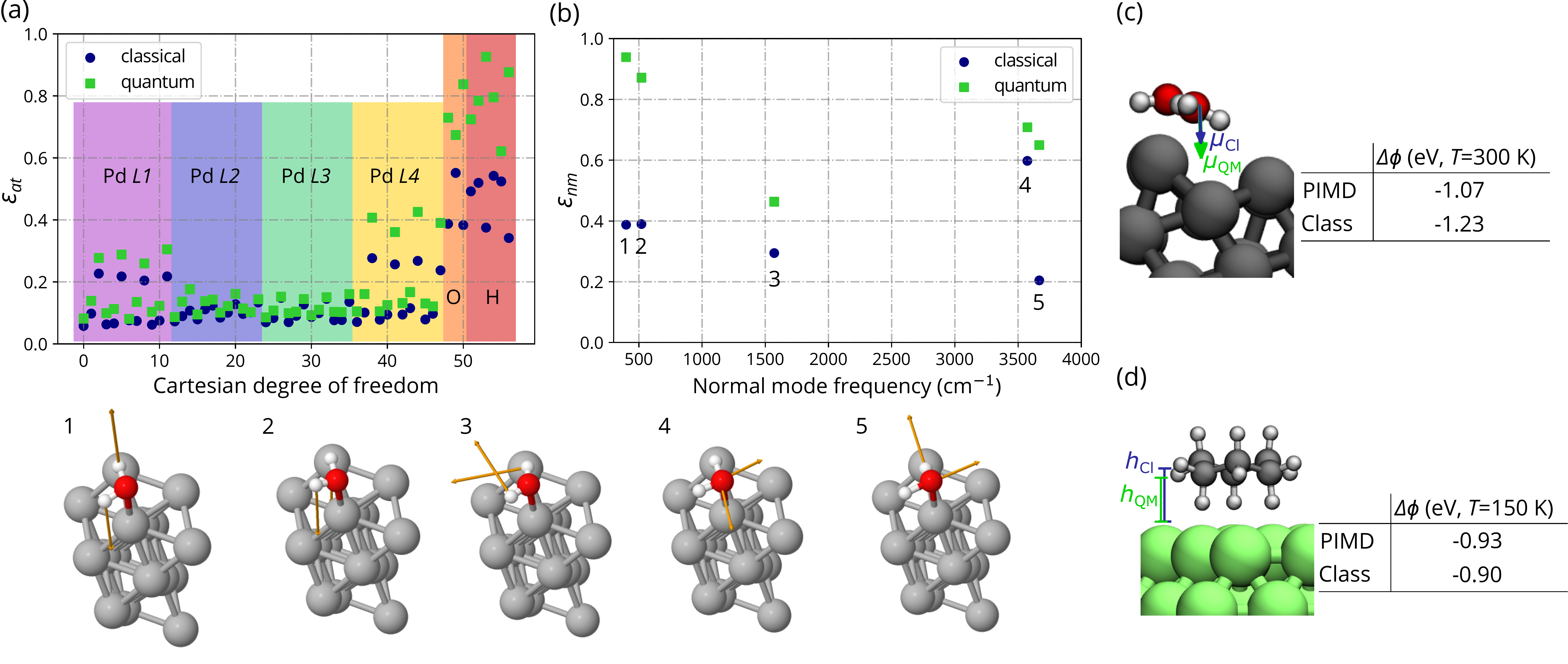}
    \caption{(a) Estimation of anhamonic contributions to forces according to the quantity $\epsilon$ defined in the text for all Cartesian degrees of freedom of a water molecule adsorbed on a Pd(111) surface with four layers, ordered as $x$, $y$, $z$ for each atom. The layers are labeled with different colors, and $L4$ is in contact with the molecule. Dark blue circles indicate $\epsilon$ calculated from classical Boltzmann statistical sampling while green squares were calculated from quantum Boltzmann statistical sampling for the degrees of freedom. (b) Same quantities as in (a) but on the normal mode representation, showing only the modes located mostly on the adsorbate, pictured at the bottom. (c) Change in the work function (DFT-PBE+vdW$^{\text{surf}}$) of water molecules adsorbed on the steps of a Pt(221) surface, as shown in Ref.~\cite{Litman:2017eh}. The molecular dipole component in the direction of the surface is larger if quantum statistics for the nuclei are considered, which causes a different work function change. (d) Change in the work function (DFT-PBE+vdW$^{\text{surf}}$) of cyclohexane adsorbed on Rh(111), as shown in Ref.~\cite{FidanyanRossi2020}. The molecule lies slightly closer to the surface if quantum statistics for the nuclei are considered, which causes a difference in work function change.}
    \label{fig:fig3}
\end{figure*}

The considerations discussed in the previous section make it clear that there is a sort of 
``competition'' involving anharmonicity and NQE, which is \rev{depicted in Fig. 
\ref{fig:fig2-next} using two models as examples}. \rev{In Fig. \ref{fig:fig2-next}a, the anharmonic contributions to the fluctuations of nuclear positions $\langle q^2 \rangle$ calculated from classical or quantum mechanics for a model where the anharmonicity is comparable across the whole frequency range are plotted. These anharmonic contributions are similar at lower frequencies and quite different at higher frequencies in this case. The frequency range for which the classical and quantum anharmonic contributions coincide increases with increasing temperature. Furthermore, while in the classical case the anharmonic contributions to these fluctuations are very small across all frequencies at low temperatures, in the quantum case these contributions are large for higher frequencies at all temperatures.  However, as shown in Fig.~\ref{fig:fig2-next}b, in a model where low-frequency vibrational modes present a higher degree of anharmonicity than higher frequency ones (which is often observed in real materials, as exemplified below), there is only a finite region at intermediate frequency ranges and intermediate to lower temperatures where quantum anharmonic contributions significantly differ from the classical estimate \textit{and} are, at the same time, of a significant magnitude (because at higher frequencies they approach zero for both classical and quantum nuclei).} 
Systems where both anharmonicity and NQE play a significant role at temperatures of interest typically involve light atoms moving on a potential energy surface in which the relevant coordinates are highly anharmonic (e.g., \rev{double-well} potentials). 

For static properties of materials, it is indeed often found that even when anharmonic effects are pronounced, they can be well-captured by classical statistical sampling of the nuclear potential energy surface, and the quantum-nuclear contributions, although important, can be well described within the harmonic approximation. In fact, for situations where it is possible to define a well-suited reference configuration, one can systematically break down the different contributions to a fully anharmonic quantum mechanical quantity into classical and quantum components, as well as harmonic and anharmonic contributions to each.
An example of such a breakdown for free energies of weakly bonded molecular crystals has been discussed in Ref.~\cite{RossiCeriotti_PRL2016}. A conclusion from that work, which has also been observed in other materials like metal-organic frameworks and ice~\cite{Kapil:2019hh,Ramirez:2012cu}, is that the largest contributions to anharmonic motions stem from low energy vibrational modes, which can be treated classically at relevant temperature ranges. This observation resonates with what has been recently reported in a study involving several solid state materials, including inorganic perovskites~\cite{KnoopCarbogno_unpub2020}. \rev{Therefore, for example in the estimation of free energy differences of molecular crystals, only a small error was observed by treating the anharmonic contributions to this quantity classically and the quantum contributions within the harmonic approximation~\cite{RossiCeriotti_PRL2016, Kapil:2019ch}}.

However, there are situations where the harmonic approximation for NQE is truly insufficient, and going beyond this approximation unravels important effects also in the electronic structure. One such situation involves weakly bonded interfaces in which light atoms like hydrogen play a prominent role. This is exemplified in Figure~\ref{fig:fig3}a for one water molecule adsorbed on a 2$\times$2 Pd(111) surface with 4 layers (FHI-aims program \textit{light} settings~\cite{Blum:2009vp}, 6$\times$6$\times$1 k-points, 100 \AA~vacuum). The anharmonic contributions to the forces of this system was estimated following the procedure presented in Ref.~\cite{KnoopCarbogno_unpub2020}. In short, the atomic displacements were sampled according to both classical and quantum Boltzmann statistics for the nuclei at $T$=100 K, with a harmonic potential defined by the Hessian matrix of the minimum-energy structure, as obtained with DFT employing the PBE functional including the vdW interactions of Ref.~\cite{Ruiz:2012gd}. Subsequently, the root mean square deviation $\Delta_i=\sqrt{\sum_s^{S} (F_{s,i}^{\text{DFT}}- F_{s,i}^{\text{h}})^2/S}$ was calculated, where $F_{s,i}^{\text{DFT(h)}}$ is the force of degree of freedom $i$ in sample $s$ calculated from the DFT (harmonic) potential. A sample size of $S=100$ was considered. For ease of comparison, $\Delta_i$ obtained with quantum and classical nuclei were normalized by the standard deviation $\sigma_i$ of the force of the corresponding degree of freedom as sampled from quantum Boltzmann statistics. We define $\epsilon_i=\Delta_i/\sigma_i$ as a measure for comparison. The result of this calculation for all the Cartesian degrees of freedom of this system, as well as for the normal modes of the adsorbate, is presented in Fig.~\ref{fig:fig3}(a) and (b). Anharmonic contributions are larger for the degrees of freedom of the adsorbate and the top/bottom layers of the slab. Interestingly, the $z$ components of the top Pd(111) layer and all the coordinates of the adsorbate show a significantly higher degree of anharmonicity when the nuclei are treated as quantum particles. Analyzing the same quantities in the normal mode representation reveals that the most anharmonic modes belonging to the adsorbate also show a pronounced contribution from anharmonic NQE, in particular for the modes that bend the hydrogen atoms towards the surface.

\begin{figure*}[ht]
    \centering
    \includegraphics[width=0.95\textwidth]{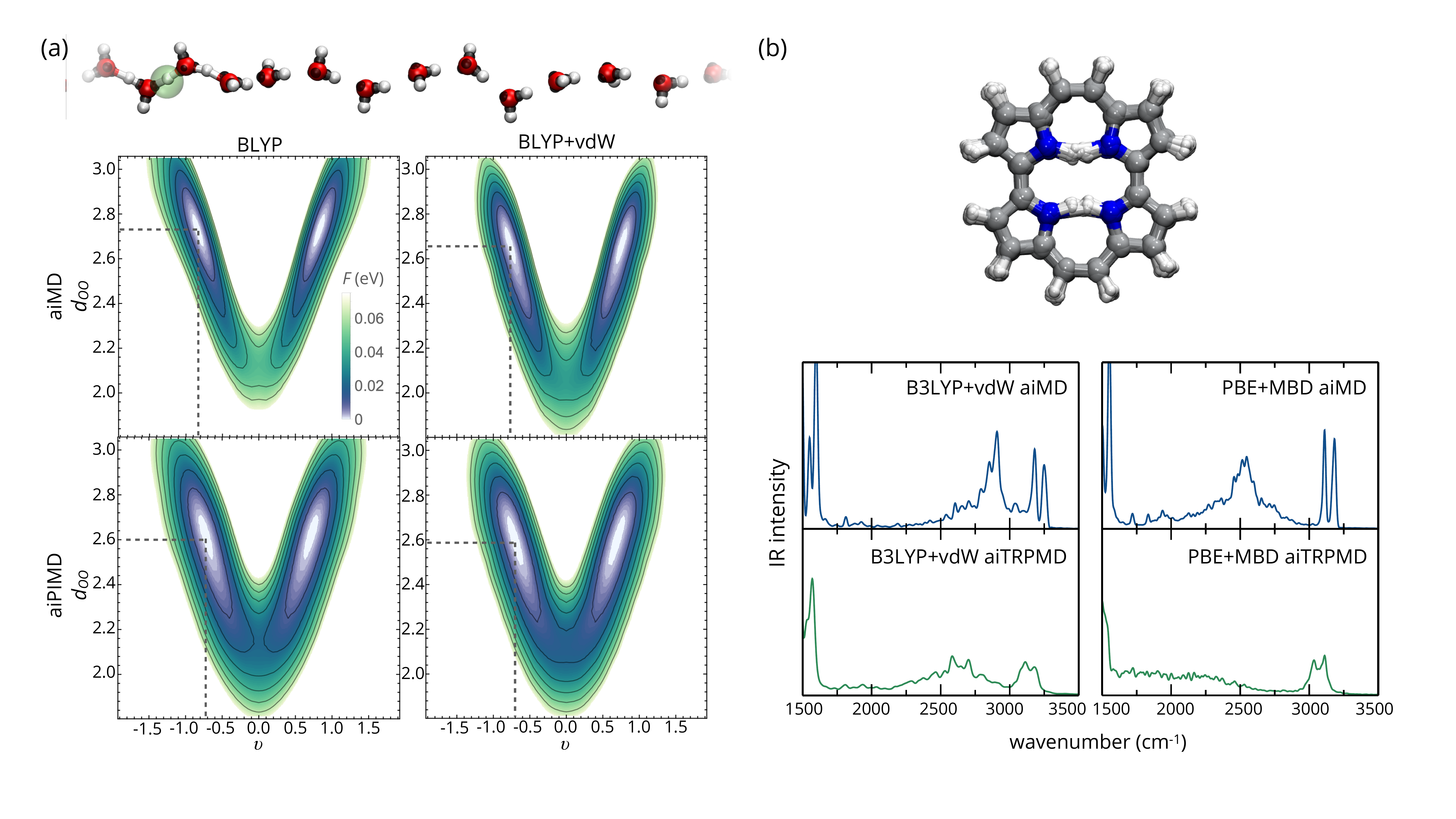}
    \caption{(a) Analysis of the dynamics of diffusion of a proton in a water wire confined by a cylindrical potential, as presented in Ref.~\cite{RossiDiff2016}. The centers of the Boys orbitals of each water monomer as calculated from DFT-BLYP(+vdW), shown in black at the top, were used as descriptors of a machine-learning clustering algorithm to identify the position of the proton at any point in time (green translucent sphere at the top). Free energy surfaces projected on the oxygen-oxygen distances $d_{\text{OO}}$ and the proton transfer coordinate $\nu = d_{\text{OH1}}-d_{\text{OH2}}$, as calculated from MD and PIMD simulations with the BLYP and the BLYP+vdW functionals. This data was presented in Ref.~\cite{RossiDiff2016}. (b) Infrared spectrum of the porphycene molecule, as calculated from MD and TRPMD calculations ($T = 300$ K) on potential energy surfaces calculated with the B3LYP+vdW functional and the PBE+MBD functional. Overlaid snapshots of the B3LYP+vdW TRPMD simulation are shown in the upper panel. This data was presented in Ref.~\cite{LitmanJACS2019}.}
    \label{fig:fig4}
\end{figure*}

These anharmonic NQE can affect the electronic structure of these interfaces -- an effect that can only be captured when both \rev{the electronic and nuclear degrees of freedom are treated within quantum mechanics}. Another striking example of these effects can be seen in a related system of water molecules adsorbed on the steps of a Pt(221) slab. The work function change $\Delta \phi$ of the interface was calculated with quantum and classical nuclei~\cite{Litman:2017eh} by performing \textit{ab initio} path integral molecular dynamics using an acceleration scheme that reduces the necessary amount of PIMD replicas in weakly bonded interfaces,  proposed in Ref.~\cite{Litman:2017eh}. As reported in that work, a significant difference in $\Delta \phi$ is introduced when considering quantum nuclei, as reproduced in Fig~\ref{fig:fig3}(b). The origin of this effect was explained to be the increase in magnitude of the molecular dipole component pointing towards the surface upon the inclusion of NQEs. 

In contrast, for a largely non-polar molecule, namely cyclohexane adsorbed on Rh(111), NQE were shown to impact the $\Delta \phi$ both in experiment and theory~\cite{Koitaya:2014gd,FidanyanRossi2020}. However, in this case the origin of the effect does not lie on anharmonic NQE on the internal degrees of freedom of the adsorbate. As depicted in Fig.~\ref{fig:fig3}(c), the anharmonicity on the molecule-surface interaction coordinate, which includes an H-Rh bond, causes the adsorbate to lie at different distances from the surface if nuclei are classical or quantum. This changes the effective surface dipole through the modulation of the density overlap of adsorbate and surface and the pushback effect. Therefore, a variety of consequences of the interplay between anharmonic NQE and electronic structure can emerge at interfaces.

The potential energy surface itself plays a decisive role in how NQE affect the structure and dynamics of any given system. One well-known and rather extreme example is that of the competing NQEs in liquid water~\cite{Habershon_2009}. When analysing the impact of NQE in dynamical properties of water with PIMD-based simulations, Habershon, Markland and Manolopoulos noticed that they were substantially overestimated when calculated on potential energy surfaces where individual water monomers were considered to be rigid, or where the internal degrees of freedom were described by simple harmonic terms. The evaluation of self-diffusion coefficients with fully flexible (and anharmonic) water molecules showed that the impact of NQE on this quantity is small. However, this is not because NQE in water are negligible. Instead, NQE in directions perpendicular to the hydrogen bond (H-bond) between molecules make the network weaker, contributing to an increase in the diffusion coefficient -- but this is compensated by NQE acting parallel to the H-bonds, which increase the molecular dipole and strengthen intermolecular interactions, causing a decrease in the diffusion coefficient. \textit{Ab initio} potential energy surfaces do not contain such drastic approximations on any degree of freedom, but they can still considerably change the impact of NQE on certain properties. For liquid water, excellent agreement with experiment for diffusion coefficients and vibrational properties have been reported~\cite{Marsalek:2017kn} when employing TRPMD with the revPBE0 functional, and these simulations also confirmed the competing NQE picture. These competing effects have also been observed in other complex H-bonded materials, like biomolecules~\cite{Rossi:2015poly,Fang:2016dna}.

Further examples where changing the \textit{ab initio} potential can provoke substantial changes of the impact of NQE on 
dynamical properties of different systems is shown in Fig.~\ref{fig:fig4}. In Ref.~\cite{RossiDiff2016} the diffusion of proton and hydroxide ions was studied with \textit{ab initio} MD and TRPMD. One advantage of using \textit{ab initio} potentials is that it was possible to define an unbiased descriptor for the position of the proton at any given time in the simulation, based on the position of the center of the Boys orbitals and an associated machine-learning clustering algorithm~\cite{RossiDiff2016}. A depiction of the location of the proton in one simulation snapshot is shown on the top of  Fig.~\ref{fig:fig4}a. 
Using this technique, it was observed that the diffusion coefficient of a proton in a water wire at 300 K is almost doubled when considering NQE on a potential energy surface that does not contain long-range van der Waals interactions (DFT-BLYP). However, the diffusion remains basically unchanged when NQE are included on a potential energy surface that contains these interactions~\cite{Tkatchenko:2009TS} (DFT-BLYP+vdW), which are known to be necessary for the description of H-bonded systems~\cite{Marom:2011vdW}. The underlying reason is depicted in the free energy plots in Fig.~\ref{fig:fig4}(a), as also discussed in Ref.~\cite{RossiDiff2016}. The oxygen-oxygen distance $d_{OO}$ in such wires is smaller when including vdW interactions, such that the barrier at the proton transfer coordinate $\nu$ is similar to $k_BT$ at room temperature, even without including NQE. The hydrogen transfer is, thus, not the rate limiting step for proton diffusion at this temperature, which lies instead on the rearrangement of the water molecules in the wire. When disregarding vdW interactions, the hydrogen transfer barrier is higher than $k_BT$ at room temperature for classical nuclei and likely becomes the rate limiting step. However, when considering quantum nuclei, zero-point energy makes the effective barrier much smaller and this ceases to be the case. 

A final example is that of the vibrational spectrum of porphycene, presented in Ref.~\cite{LitmanJACS2019} and depicted in Fig. \ref{fig:fig4}b. The internal hydrogen atoms in this molecule exhibit a strongly anharmonic potential energy profile along the hydrogen-transfer coordinate, characterized by a double-well. Among several DFT functionals considered in Ref.~\cite{LitmanJACS2019}, B3LYP with pairwise vdW corrections (B3LYP+vdW) was the one that best approximated the relative energies of the stationary points in the potential energy surface calculated with coupled-cluster including single, double and perturbative triple excitations (CCSD(T)). The calculation of the anharmonic infrared (IR) spectrum of porphycene with this functional shows that the NH-stretch signal, located in the region between 2000 and 3000 cm$^{-1}$, is broadened and red-shifted when approximating NQEs with TRPMD. The TRPMD result matches experiment very well~\cite{LitmanJACS2019} and a strengthening of the H-bonds was observed upon including NQE. In contrast, the PBE functional with many-body dispersion interactions~\cite{Tkatchenko:2012MBD,DiStasioJr:2014MBD} (MBD) predicts much lower barriers for the hydrogen transfer in this system, compared to CCSD(T). In fact, they are so low that a harmonic estimate of ZPE showed that they would completely ``fill'' the double well, making the barrier irrelevant, as discussed in Ref.~\cite{LitmanJACS2019}. Indeed, when calculating the IR spectrum with this functional from classical-nuclei MD, the NH-stretch peak is already considerably red-shifted (almost centered at the position that B3LYP-TRPMD predicts). When including NQE, it becomes extremely broad and overly red-shifted, in complete disagreement with experiments. Anharmonic effects in porphycene are so pronounced that intermode coupling leads to a qualitatively different temperature dependence on the position of this same peak if nuclei are considered as quantum or classical particles~\cite{Litman:2019ix}. This phenomenon is bound to be observed in a wide range of high-dimensional anharmonic systems with significant coupling between high and low frequency vibrational modes, which can now be addressed with appropriate methodology. \rev{It is also coupling between vibrational modes of surface and adsorbate that results in a surprising tunneling rate enhancement of intramolecular hydrogen transfer when porphycene is adsorbed on metallic surfaces~\cite{LitmanRossi2020}}.

\section{Final Remarks}

Including NQE beyond the harmonic approximation in \textit{ab initio} simulations is within reach in many applications in chemical physics, condensed matter physics and materials science. As these areas become more focused on soft organic materials for technological applications, these methods will undoubtedly deliver the much needed and unprecedented fundamental insights into many problems where both electrons and nuclei require a quantum-mechanical description. They are also relevant for understanding matter and chemistry in extreme conditions as found, e.g., in interstellar space or the interior of planets.

At the same time, the recent methodological developments discussed here have pushed the boundaries where the real challenges in these simulations lie. In this perspective, a few methodological challenges were mentioned, namely, the need for practical and more accurate methods that can capture dynamical NQE, and the need of overcoming the barrier of performing these simulations with more accurate electronic structure methods. The former still seems slightly elusive, but the field is quite active and there is hope for new solutions. The latter, on the other hand, can be addressed by taking advantage of the recent developments in machine-learning techniques for atomistic simulations, coupled with the possibility of reusing data stored in large databases.

An important fundamental next step is the development of practical methodology that is able to capture electron-phonon coupling in non-adiabatic scenarios and where many anharmonic (quantum) nuclear degrees of freedom play an important role -- possibly also in non-equilibrium situations. The open questions in this area are considerably more numerous, but will hopefully also profit from the advances achieved so far for the adiabatic regime.

\section*{Acknowledgements}

I would like to acknowledge the members of my research group, who have greatly inspired me throughout the past years. In the spirit of this special issue, I would also like to thank the numerous mentors, collaborators, and especially female mentors, collaborators, and colleagues who work in my area. All of them have given me strength and inspiration to continue on this path. \rev{I also thank Michele Ceriotti and Aaron Kelly for reading earlier versions of this perspective and for very useful discussions.}

\section*{Data Availability}
The data that support the findings of this study are available from the corresponding author upon reasonable request.

\bibliography{biblio}

\end{document}